\documentstyle[aps]{revtex}

\begin{document}
\title{Secure direct communication using step-split Einstein-Podolsky-Rosen pair}
\author{Qing-yu Cai}
\address{Wuhan Institute of Physics and Mathematics, Chinese Academy of Science,\\
Wuhan 430071, People's Republic of China}
\maketitle

\begin{abstract}
We presen a secure direct communication protocol by using step-split
Einstein-Podolsky-Rosen (EPR) pair. In this communication protocol, Alice
first sends one qubit of an EPR pair to Bob. Bob sends a receipt signal to
Alice through public channel when he receives Alice's first qubit. Alice
performs her encoding operations on the second qubit and sends this qubit to
Bob. Bob performs a Bell-basis measurement to draw Alice's information. The
security of this protocol is based on `High fidelity implies low entropy'.
If Eve want to eavesdrop Alice's information, she has to attack both qubits
of the EPR pair, which results in that any effective eavesdropping attack
can be detected. Bob's receipt signal can protect this protocol against the
eavesdropping hiding in the quantum channel losses. And this protocol is
strictly secure to perform a quantum key distribution by using
Calderbank-shor-steane codes.
\end{abstract}


Quantum cryptography (QC) or quantum key distribution (QKD) exploits the
principles of quantum mechanics to enable secure distribution of private
information. In a QKD protocol, if the sender (Alice) wishes to send message
to receiver (Bob), Alice should first encode her message on the travel
qubit(s) and Bob should perform a decoding measurement on the travel
qubit(s) to gain Alice's message. To perfectly complete the QKD, a reliable
public channel is needed. The security of the resulting key is guaranteed by
the properties of quantum information, and thus is conditioned only on
fundamental laws of physics being correct. Since Bennett and Brassard
proposed the first QKD protocol in 1984 [1], a lot of theoretical QKD
protocol have been advanced [1-12]. The aim of these QKD protocols to
establish a common random key between two parties. Beige $et$ $al$. first
proposed a quantum secure direct communication (QSDC) scheme [13] in which
the message can be read only after a transmission of an additional classical
information for each qubit. In Ref.[14], Bostr\"{o}m and Felbinger presented
a ping-pong protocol which allows the information to be transmitted in a
direct way, i.e., which does not needs to establish a random key to encrypt
messages. And the ping-pong protocol has been extended by using single
photon in a mixed state [15]. In this paper, we show a communication
protocol with a strict security proof, which allows information transferred
in two-step and one local operation encoding two bit information by using
step-split EPR pair. Actually, the basic idea that one bit information can
be transferred in two-step was presented in Ref.[6]. Two localized wave
packets encoded one bit information are sent from Alice to Bob along two
separated channels. And that one local operation can encode two bits
information is well known as dense encoding [5].

Let us first start with a brief description of a secure direct communication
presented by Deng $et$ $al.$ recently [16]. Alice prepares an order N EPR
pair in the state $|\psi ^{-}>$. Alice sends half of each EPR pair to Bob.
Alice and Bob randomly select some of the EPR pairs to check Eve's
eavesdropping and other ERP pairs to transfer information. When Alice
believes that the communication is secure, she performs a unitary operation
on the qubit she kept to encode two classical bits and sends this qubit to
Bob. Bob performs a Bell-basis measurement to decode Alice's information.
Unfortunately, there are many loopholes in there secrrity proof. Eve can not
gain any of Alcie's information if she only attack the first qubit. Second,
equation (21) is wrong because that the basis $\{|0,\varepsilon
_{00}>,|1,\varepsilon _{01}>,|1,\varepsilon _{00}>,|0,\varepsilon _{01}>\}$
are not mutual orthogonal in general. So the resulting conclusion is
unreliable. As is well explained in the ping-pong protocol [14], dense
coding feature has to be abandoned in favor of a secure transmission on this
occasion ( Actually, the capacity of the ping-pong protocol can be improved
doubly with another security proof. [17]).

As is well known, an ERP pair can be in one of the four Bell states, 
\begin{eqnarray}
|\phi ^{+} &>&=\frac{1}{\sqrt{2}}(|0>|0>+|1>|1>)  \nonumber \\
&=&\frac{1}{\sqrt{2}}(|+>|+>+|->|->),
\end{eqnarray}
\begin{eqnarray}
|\phi ^{-} &>&=\frac{1}{\sqrt{2}}(|0>|0>-|1>|1>)  \nonumber \\
&=&\frac{1}{\sqrt{2}}(|+>|->+|->|+>),
\end{eqnarray}
\begin{eqnarray}
|\psi ^{+} &>&=\frac{1}{\sqrt{2}}(|0>|1>+|1>|0>)  \nonumber \\
&=&\frac{1}{\sqrt{2}}(|+>|+>-|->|->),
\end{eqnarray}
\begin{eqnarray}
|\psi ^{-} &>&=\frac{1}{\sqrt{2}}(|0>|1>-|1>|0>)  \nonumber \\
&=&\frac{1}{\sqrt{2}}(|+>|->-|->|+>),
\end{eqnarray}
where $|+>=\frac{1}{\sqrt{2}}(|0>+|1>)$, $|->=\frac{1}{\sqrt{2}}(|0>-|1>)$,
and $|0>$, $|1>$ are the up and down eigenstate of the $\sigma _{z}$. It is
well known that such EPR pairs can be used to establish nonlocal
correlations over a spacelike interval. But these correlations cannot be
used for superluminal communication. Consider Alice has an EPR pair in the
singlet state $|\psi ^{-}>$. She sends one qubit to Bob and keeps another.
Assume both Alice and Bob perform measurement on their qubits in basis $%
B_{z} $, $B_{z}=\{|0>$, $|1>\}$. When Alice's measurement outcome is $|0>$,
then she immediately knows Bob's measurement outcome is $|1>$. If Alice's
measurement outcome is $|1>$, then she knows Bob's measurement outcome is $%
|0>$. But whether Alice's measurement outcome is $|0>$ or $|1>$ is
completely random with a probability $p=0.5$. As long as one of them
performs a measurement, the state $|\psi ^{-}>$ will instantaneously
collapse to a product state $|0>_{A}|1>_{B}$or $|1>_{A}|0>_{B}$ randomly.
The correlations do not exist any longer. So Bob can not get any of Alice's
message through these processes. In order to sends message from Alice to
Bob, a reliable public channel is needed.

When two qubits (A and B) are in the maximally entangled states $|\psi ^{\pm
}>$ and $|\phi ^{\pm }>$, each single qubit is in the completely mixed
states, $\rho _{A}^{\pm }:=tr_{B}\{|\psi ^{\pm }><\psi ^{\pm
}|\}=tr_{B}\{|\phi ^{\pm }><\phi ^{\pm }|\}$. No one can distinguish these
states by an experiment performed on only one qubit. In other words, one
qubit can be encoded by a local operation but it has a nonlocal effect.
Anyone who only access to one of the qubits can not decode the information
if he has no access to the other qubit. Suppose Alice has an EPR pair in the
state $|\psi ^{-}>$. She sends one qubit (we call it `the first qubit') to
Bob and keeps another. When Bob receives the first qubit, he sends a signal
to Alice through public channel (we call this as Bob's $receipt$). If Alice
receives Bob's receipt, she then performs an encoding operation $U$ on her
qubit and then sends this qubit (we call this qubit as `the second qubit')
to Bob. Bob performs a Bell-basis measurement on two qubits to decode
Alice's information. Alice's encoding operation $U$ can be described by 
\begin{eqnarray}
U_{00} &=&\left( 
\begin{array}{ll}
1 & 0 \\ 
0 & 1
\end{array}
\right) ,U_{01}=\left( 
\begin{array}{ll}
1 & 0 \\ 
0 & -1
\end{array}
\right)   \nonumber \\
U_{10} &=&\left( 
\begin{array}{ll}
0 & 1 \\ 
1 & 0
\end{array}
\right) ,U_{11}=\left( 
\begin{array}{ll}
0 & 1 \\ 
-1 & 0
\end{array}
\right) .
\end{eqnarray}
These four operations can transform the state $|\psi ^{-}>$ into $|\psi ^{-}>
$, $|\psi ^{+}>$, $|\phi ^{-}>$, and $|\phi ^{+}>$, respectively. These
operations respectively correspond to the code 00, 01, 10, 11. To ensure the
security of this communication, a check mode is needed. When Alice receives
Bob's receipt, instead of her encoding operation, she performs a measurement
in the basis $B_{z}=\{|0>,$ $|1>\}$ or $B_{x}=\{|+>,|->)\}$ randomly. Then
she tells her measurement outcome to Bob. Bob also performs a measurement in
the same basis as Alice used. If both outcomes coincide, they known that Eve
is in line. This communication stops. Else, Alice prepares next EPR pair.
This protocol an explicit described like an algorithm.

\begin{description}
\item  (1) Alice prepares an EPR pair in state $|\psi ^{-}>$.

\item  (2) Alice sends one qubit to Bob and keeps another.

\item  (3) Bob receives the first qubit and sends a classical signal to
Alice.

\item  (4) When Alice receives Bob's receipt, she select encoding operation
(5e) or checking measurement (5c) randomly.

\item  (5c) Alice performs a measurement in basis $B_{z}$ or $B_{x}$
randomly. She tells Bob her measurement result to Bob through public
channel. Bob also performs a measurement in the basis Alice used. If both
measurement outcomes coincide, there is Eve in line. This communication
stops. Else, this communication continues.

\item  (5e) Alice performs an encoding operation. She sends the second qubit
to Bob. Bob receives the second qubit. He performs a Bell-basis measurement
on two qubits to decode Alice's information.

\item  (6) When all of Alice's information is transmitted, Alice and Bob
announce some codes as message authentification.

\item  (7) This communication successfully terminated.
\end{description}

$\ Security$ $proof$. Since $\rho _{A}^{\pm }:=tr_{B}\{|\psi ^{\pm }><\psi
^{\pm }|\}=tr_{B}\{|\phi ^{\pm }><\phi ^{\pm }|\}$, Eve can not distinguish
each Bell state if she only attack one qubit of the EPR pair, i.e., Eve can
not get any information if she only attack the second qubit. Because Alice
decides to perform check measurement only after Bob received the first
qubit, the security proof has to assure Eve's any effective eavesdropping
attack can be detected. To gain Alice's information, Eve has to attack both
qubits of the EPR pair. Suppose after Eve's first attack, the state Alice
and Bob shared becomes $\rho $. The information Eve can gain from $\rho $ is
bounded by the Holevo quantity $\chi (\rho )$ [18]. Because Holevo quantity
decreases under quantum operations [18,19], the mutual information Eve can
gain after Alice's encoding operation is determined by $\chi (\rho )$. From 
\begin{equation}
\chi (\rho )=S(\rho )-\sum_{i}p_{i}S(\rho _{i}),
\end{equation}
we know $S(\rho )$ is the upper bound of $\chi (\rho )$. `High fidelity
implies low entropy'. The fidelity [20] of state $|\psi ^{-}>$ and $\rho $
is 
\begin{equation}
F(|\psi ^{-}>,\rho )=\sqrt{<\psi ^{-}|\rho |\psi ^{-}>}.
\end{equation}
Let us assume that 
\begin{equation}
F(|\psi ^{-}>,\rho )^{2}=<\psi ^{-}|\rho |\psi ^{-}>=1-\gamma ,
\end{equation}
where $0\leq \gamma \leq 1$. Therefore, the entropy of $\rho $ is bounded
above by the entropy of a diagonal density matrix $\rho _{\max }$ with
diagonal entries $1-\gamma $, $\frac{\gamma }{3}$, $\frac{\gamma }{3}$, $%
\frac{\gamma }{3}$. And the entropy of $\rho _{\max }$ is 
\begin{equation}
S(\rho _{\max })=-(1-\gamma )\log _{2}(1-\gamma )-\gamma \log _{2}\frac{%
\gamma }{3}\text{.}
\end{equation}
So $S(\rho _{\max })$ is a upper bound of the information Eve can gain from $%
\rho $. Let us discuss the connection between the information Eve can gain
and the detection probability $d$. In check mode, when Alice and Bob share
the state $|\phi ^{\pm }>$, their measurement outcomes will coincide every
time when they use the measurement basis $B_{z}$. When the state they shared
is $|\psi ^{+}>$, their measurement outcomes will coincide every time when
they use the measurement basis $B_{x}$. If and only if the state they shared
is $|\psi ^{-}>$, their measurement results will never coincide. Since $%
F(|\psi ^{-}>,\rho )^{2}=1-\gamma $, then the detection probability is $%
d\geq \gamma /2$. From equation (9) we know when $\gamma =0$, i.e., Eve does
not eavesdrop the communication, the detection probability $d=0$. When $%
\gamma >0$, i.e., Eve can gain some of Alice's information, she has to face
a nonzero risk $d>0$ to be detected. When $\gamma =\frac{3}{4}$, it has $%
S(\rho _{\max })=2$, which implies that Eve has a chance to eavesdrop full
of Alice's information. In fact, Eve can reach this upper bound. She
replaces Alice's first qubit with one of her own and forwards it to Bob.
When she receives the seconds qubit, she performs a Bell-basis measurement
on both qubits to draw full of Alice's information. On this occasion, the
detection rate is $d\geq \frac{3}{8}$.

Eve can attack the communication without eavesdropping [21]. She only attack
the second qubit to destroy the communication. In this case, Eve can not
gain Alice's information. After Alice's encoding operation, Eve captures the
second qubit and performs a measurement in the basis $B_{x}$ or $B_{z}$ on
it, which makes the state of the EPR pair collapse to a product state. Eve
also can perform a unitary operation on the second qubit. She can performs
the operation $U_{01}$, $U_{10}$, or $U_{11}$ on every second qubit to
attack the communication. Then Eve forwards the seconds qubit to Bob. When
the communication is terminated, Bob has learned nothing but a sequence of
nonsense random bits. Alice and Bob can use a message authentification
method to protect the communication against Eve's denial-of-service attack
with a reliable public channel [21].

Consider that the quantum channel is noisy. Noise in quantum channel will
introduce qubit losses and qubit errors. In a noisy quantum channel, Eve's
eavesdropping may be hidden in the quantum channel losses. In our
communication protocol, Bob sends a receipt in every run. Without this
signal, quantum channel losses may be used by Eve to eavesdrop Alice's
information [22]. Eve can keeps the first qubit for a time. If Alice publish
her measurement outcome, Eve then forwards the first qubit to Bob. Else, Eve
receives the second qubit and performs a Bell-basis measurement to draw
Alice's information. Thus, such eavesdropping can be hidden in the quantum
channel losses without Bob's receipt signal. Alice and Bob can use the
message authentification method to detect Eve's eavesdropping hidden in
qubit error. If Alice and Bob find the error rate is higher than they
desired, they abandon this communication. With a low noise channel, Alice
and Bob can use Calderbank-Shor-Steane (CSS) codes [23] to complete a
perfect secure communication. In this way, our protocol is similar to the
modified Lo-Chau protocol [24]. It is strictly secure to complete a key
distribution.

In summary, we present a secure direct quantum communication based on
step-split EPR pair. In principle, this protocol allows quasisecure direct
communication and secure QKD. In practice, the storage of one photon is
necessary for a duration corresponding to the distance between Alice and Bob
[25]. Today, the Bell state of an EPR photons can be created by parametric $%
down-conversion$. And the complete Bell type measurement is also been
demonstrated [26]. The values of $\sigma _{x}$, $\sigma _{y}$, and $\sigma
_{z}$ of a qubit of an single photon can be ascertained [27]. With current
technologies, the experimental realization of the protocol is feasible with
relative small effort.

\section{acknowledgement}

This work is funded by National Nature Science Foundation of China (Grant
No. 10374119).

\section{reference}

[1] C. H. Bennett, and G. Brassard, in Proceedings of the IEEE international
Conference on Computers, Systems and Signal Processing, Bangalore, India
(IEEE, New York, 1984), pp. 175-179.

[2] A. K. Ekert, Phys. Rev. Lett. 67, 661 (1991).

[3] C. H. Bennett, G. Brassard, and N. D. Mermin, Phys. Rev. Lett. 68, 557
(1992).

[4] C. H. Bennett, Phys. Rev. Lett. 68, 3121 (1992).

[5] C. H. Bennett and S. J. Wiesner, Phys. Rev. Lett. 69, 2881 (1992).

[6] L. Goldenberg and L. Vaidman, Phys. Rev. Lett. 75, 1239 (1995).

[7] B. Huttner, N. Imoto, N. Gisin, and T. Mor, Phys. Rev. A 51, 1863 (1995).

[8] M. Koashi and N. Imoto, Phys. Rev. Lett. 79, 2383 (19970.

[9] D. Bruss, Phys. Rev. Lett. 81, 3018 (1998).

[10] W. Y. Hwang, I. G. Koh, and Y. D. Han, Phys. Lett. A 244, 489 (1998).

[11] H.-K. Lo and H. F. Chau, Science 283, 2050 (1999).

[12] A. Cabell, Phys. Rev. Lett. 85, 5635 (2000).

[13] A. Beige, B.-G. Englert, C. Kurtsiefer, and H. Weinfurter, Acta. Phys.
Pol. A 101, 357 (2002).

[14] K. Bostr\"{o}m and T. Felbinger, Phys. Rev. Lett. 89, 187902 (2002).

[15] Qing-Yu Cai and Bai-Wen Li, Chin. Phys. Lett. 21(4), 601 (2004).

[16] Fu-Guo Deng, Gui Lu Long, and Xiao-Shu Liu, Phys. Rev A 68, 042317
(2003).

[17] Qing--yu Cai and Bai-wen Li, Phys. Rev. A (to be published).

[18] M. A. Nielsen and I. L. Chuang, Quantum computation and Quantum
Information (Cambridge University Press, Cambridge, UK, 2000).

[19] Qing-yu Cai, arXiv:quant-ph/0303117 (unpublished).

[20] C. A. Fuchs, arXiv: quant-ph/9601020; H. Barnum, C. M. Caves, C. A.
Fuchs, R. Jozsa, and B. Schumacher, Phys. Rev. Lett. 76, 2818 (1996).

[21] Qing-yu Cai, Phys. Rev. Lett. 91, 109801 (2003).

[22] A. W\'{o}jcik, Phys. Rev. Lett. 90, 157901 (2003).

[23] A. R. Calderbank and P. W. Shor, Phys. Rev. A 54,1098 (1996);A. M.
Steane, Proc. R. Soc. London A 425, 2551 (1996).

[24] P. W. Shor and J. Preskill, Phys. Rev. Lett. 85, 441 (2000).

[25] M. F. Yanik and S. Fan, Phys. Rev. Lett. 92, 083901 (2004).

[26] Y.-H. Kim, S. P. Kulik, and Y. Shih, Phys. Rev. Lett. 86, 1370 (2001).

[27] L. Vaidman, Y. Aharonov, and D. Z. Albert, Phys. Rev. Lett. 58, 1385
(1987); O. Schulz, R. Steinhuebl, M. Weber, B.-G. Englert, C. Kurtsiefer,
and H. Weinfurter, Phys. Rev. Lett. 90, 177901 (2003).

\end{document}